\documentclass{aa}
\usepackage{graphicx}
\usepackage{txfonts}
\begin{document}

\title{Seismic constraints on open clusters} 

\author{L. Piau\inst{1}, J. Ballot\inst{2} and S. Turck-Chi\`eze\inst{2}} 

\institute{Institut d'Astronomie et d'Astrophysique, ULB, CP226, 1050
           Brussels, Belgium\\
\email{piau@astro.ulb.ac.be}
\and
           CEA/DSM/DAPNIA/Service d'Astrophysique, CE Saclay,
           91191 Gif-sur-Yvette Cedex 01, France\\
\email{jballot@cea.fr, cturck@cea.fr}
}

\abstract{
The aim of this theoretical and modelling 
paper is to derive knowledge on the global and structural parameters
of low-mass stars
using asteroseismology and taking advantage
of the stellar collective behavior within open clusters.
We build stellar models and compute the seismic signal expected
from main sequence objects in the $0.8-1.6 M_{\sun}$ range. We first
evaluate apparent magnitudes and oscillations-induced luminosity 
fluctuations expected in the Hyades, the Pleiades and the $\alpha$ Persei clusters. 
The closest cluster presents a feasible challenge to
observational asteroseismology in the present and near future. The
remainder of the work therefore focuses on the Hyades. 

We combine seismological and classical computations to address three questions:
what can be inferred about 1) mass, 2) composition and 3) extension 
of outer convection zones of solar analogs in the Hyades. 
The first issue relies on the strong sensitivity of the
large separation to mass. We show that seismic constraints
provide masses to a precision level ($0.05 M_{\sun}$) that is competitive
with the actual mass estimations from binary systems.
Then large separations ($\Delta \nu$) and second differences 
($\delta_2 \nu$) are used to 
respectively constrain metal and helium fractions in the
Hyades.
When plotted for several masses, the relation of 
effective temperature ($T_\mathrm{eff}$) 
vs large separation ($\Delta \nu$) is found to be
strongly dependent on the metal content. Besides 
this the second difference main
modulation is related to the second ionization of helium.
An accuracy in the helium mass fraction of 0.02 to 0.01
can be achieved provided mass and age
are accurately known, which is the case for a few Hyades binary 
systems. The second difference
modulations are also partly due to the 
discontinuity in stellar stratification
at the convective envelope / radiative core transition. They
permit direct insight in the stellar structure. We compute 
acoustic radii of the convective bases 
for different values of the mixing length theory
parameter $\alpha_\mathrm{MLT}$ in convection modelling, i.e. different 
convective efficiency in the
superadiabatic layers. For a given effective
temperature we show that the acoustic radius changes with convection 
efficiency. This suggests that seismology can
provide constraints on the extension of outer convection and
also more generally on the direct
approaches of convection and dynamical phenomena
being currently developed.

%\keywords{Stars : low-mass, oscillations, open clusters. 
%Stars : interiors, convection.}
}

\maketitle

\section{Introduction}\label{sec1}

Helioseismology achieved many successes in allowing direct
determinations of the depth of the solar convection zone (Christensen-Dalsgaard, Gough 
\& Thomson 1991) or photospheric mass fraction of helium (Basu \& Antia 1995). In 
a comparable fashion the development of space-borne asteroseismic experiments 
within the coming years and the increasing number of ground-based observations
should yield extremely precious data 
on other stars. In this paper we discuss some aspects of the possible contribution of
asteroseismology to the determination of mass, composition and structure of 
solar-like stars.
We address the Hyades, Pleiades and $\alpha$ Persei open clusters.
These clusters have been extensively studied from theoretical and 
observational viewpoints (see Lebreton 2000 for a recent review) and their stars
are among the targets of future asteroseismic missions such as EDDINGTON
(Favata et al. 2000).
Some of them will also be interesting targets for the recent and
promising instruments HARPS (Pepe et al. 2003) or ESPaDOnS (Donati, Catala \& Landstreet 2004).  
Contrary to field stars they offer the possibility to
combine observations for stars having similar ages and
compositions. In consequence they make it possible to exploit collective effects
in seismic and stellar properties of stars with different masses. 
Finally they span a range of age and composition that should
translate into significant differences in the seismic signals of solar-like stars.

We compute stellar models and subsequently use them to predict
oscillation frequencies. We use the CESAM code (Morel 1997) for pre-main
sequence and main sequence modelling. Above 5800 K our 
version of the code relies on the 
OPAL2000 equation of state (Rogers, Swenson \& Iglesias 1996) 
and opacities (Iglesias \& Rogers 
1996). For lower temperatures the OPAL equation of state is replaced by 
the MHD equation of state (Mihalas, Dappen \& Hummer 1988) and the opacities are from 
Alexander \& Ferguson (1994). The atmosphere is restored 
using Hopf's law (Mihalas 1978). It is connected to the envelope
at optical depth 10 where the diffusion approximation for radiative 
transfer becomes valid (Morel et al. 1994). Microscopic
diffusion is always taken into account following the Michaud \& Proffitt
(1993) prescriptions.
Finally the convection zones are assumed to be fully
homogeneous, and modeled using the mixing-length theory
(hereafter MLT) in a formalism close to that of B\"ohm-Vitense (1958).
Unless mentioned explicitly 
we consider $\alpha_\mathrm{MLT}=1.766$ which is our Sun-calibrated 
value. No overshooting has been considered
in convective cores or outer convective regions.
Once the stellar structure has been obtained we use the 
adiabatic pulsation package of J{\o}rgen Christensen-Dalsgaard (1982) to
compute oscillation frequencies\footnote{Aarhus
Adiabatic Pulsation Package, presently available at 
http://astro.phys.au.dk/$\sim$jcd/adipack.n/}. 
As usual $n$
and $\ell$ denote respectively the radial order
and the degree of the mode whose frequency is written as $\nu_{\ell,n}$.

Before we present the approaches we used to
relate the stellar parameters to the
seismic signal we address the question of the modes amplitudes in the
following section. This question is a crucial one for
the efficiency of future seismic campaigns.

In the Sect. 3 we focus on the impact of
global parameters on the asteroseismic data. Both initial 
mass and chemical composition are
important in stellar evolution. Helium is the second
most abundant constituent of the Universe. It is
is however impossible to determine the helium fraction
from spectral analysis of low-mass stars.
Its abundance therefore remains undetermined in all open clusters
except the Hyades. In this open cluster lower-main sequence fits
that constrain helium content have been performed (Perryman et al. 1998; 
Pinsonneault et al. 1998). 
Stellar mass determination based on close binaries is similarly a difficult task.
Except in the case of nearby systems such as $\alpha$ Cen the
uncertainty in the mass generally is > 0.1 $M_{\sun}$ (S\"oderhjelm 1999).
Addressing the specific case of the Hyades cluster we show what
improvements seismology could give regarding the helium mass fraction
and more generally composition.

In Sect. 4  we discuss the seismic characterisation of 
the transition from the radiative interior to the convective 
envelope in solar-type stars.
Then we evaluate the impact of changes in the convective efficiency
in the superadiabatic regime over the depth of the convection zone.
We therefore propose seismology as a tool
to calibrate the $\alpha_\mathrm{MLT}$ parameter or more importantly the
convection efficiency to be inferred from direct hydrodynamic
computations for different surface conditions ($T_\mathrm{eff}$, $\log\,g$).

A discussion of our results and our conclusions
are given in Sect. 5.

\section{Detectability of oscillations}\label{sec2}

The detection and study of solar-like oscillations
modes are restricted by several factors.
Not only are they limited to
a certain range of visual apparent magnitude but they also
depend on the amplitude of the luminosity modulations.
In this section we discuss the anticipated amplitude
of the p-modes for the various stars we modeled.
This is a central point in the perspective of space asteroseismic
missions such as COROT (Baglin et al. 2002) and MOST (Walker et al. 2003)
or ground based observations.
We have derived the apparent magnitudes in V band on the 
$\sim 0.77$ to $1.6 M_{\sun}$ stellar mass range and for the
three open clusters (Table \ref{tab1}).
We take $M_{\mathrm{Bol}}=4.72-2.5\log_{10}(L/L_{\sun})$
(ie $M_{\mathrm{Bol\, \sun}}=4.72$)
and we apply the bolometric correction $BC_V$ as deduced from
a bilinear interpolation in $T_\mathrm{eff}$ and 
$\log\,g$ tables provided in 
Houdashelt et al. (2000)\footnote{IAU Commission 25 (1997) XIII 
General Assembly recommended the bolometric 
magnitude of the Sun to be defined as +4.75. However, following Houdashelt et al. (2000)
we use +4.72 for consistency.} where $T_{eff}$ is the effective temperature
and $\log\,g$ the log of the surface gravity.
We have used distance moduli based on HIPPARCOS parallaxes have been used.
We take $m-M=5.37\pm0.07$ and $6.31\pm0.08$ respectively for the Pleiades
and the $\alpha$ Persei cluster (van Leeuwen 1999).
For the Hyades we assume $m-M=3.33\pm 0.01$ (Perryman et al. 1998).
The extinction $A_\mathrm{v}$ is negligible for the Hyades, close
to 0.04 mag for the Pleiades and 0.1 mag for $\alpha$ Persei (Pinsonneault et al. 1998).
Owing to these values no reddening has been taken
into account in the process of computing visual apparent magnitudes 
from the absolute bolometric magnitudes provided by the models.
The magnitudes we report are therefore only accurate to 0.1 mag.

\begin{table*}[ht]
  \begin{center}
    \caption{Apparent visual magnitude $m_V$ and expected luminosity fluctuations
for 0.78, 0.8, 0.95, 1, 1.15, 1.3, 1.5 and 1.6 $M_{\sun}$ stellar models. Assumed ages are
90, 125 and 625 Myr for $\alpha$ Persei, Pleiades and Hyades respectively. We consider
non-solar repartition of metals in the Hyades and Y=0.26 (see Sect. \ref{sec3}).
For the two other clusters we assume solar composition. Luminosity fluctuations
are probably underestimated for the $1.5$ and $1.6 M_{\sun}$ models. Rows
4 to 7 list various seismological features for the age and composition of the Hyades.}\vspace{1em}
    \renewcommand{\arraystretch}{1.2}
    \begin{tabular}[h]{lcccccccc}
      	Mass $M_{\sun}$ & $0.78$ &  $0.8$ &  $0.95$ &  $1$ &  $1.15$ &  $1.3$ &  $1.5$ &  $1.6$ \\
      \hline
	$m_V$ (Pleiades) & 12.1   & 12.0   &    10.9   & 10.6 & 9.8     & 9.1& 8.4 & 8.1 \\
      \hline
	$m_V$ ($\alpha$ Persei) & 13.1  & 12.9  & 11.8 &  11.5 & 10.7 & 10.1 & 9.3 & 9.1 \\
      \hline
	 $m_V$ (Hyades)   & 10.1   & 9.9    &   8.8   & 8.5  & 7.7     & 7.0& 6.3 & 6.0 \\
	Luminosity fluctuations (ppm) & 1.8 & 1.9 & 2.9 & 3.2 & 4.6 & 6.2 & 8.7 & 9.9 \\
	Frequency of maximum amplitude $\nu_\mathrm{max}$ (mHz) & 3.5 & 3.5 & 3.9 & 3.9 & 3.7 & 3.3 & 2.7 & 0.8 \\
        Cut-off frequency $\nu_\mathrm{c}$ (mHz) & 8.7 & 8.5 & 6.9 & 6.3 & 4.8 & 3.6 & 2.4 & 2.1 \\
%	Hyades composition  (see text) \& age = 625 Myr &  \\
      \hline
      \end{tabular}
   \label{tab1}
  \end{center}
\end{table*}

Let us now focus on the expected amplitudes for the modes.
These are determined by the relative efficiencies of the driving and
damping processes. For solar analogs the most probable explanation of
driving of modes consist of the fluctuations of the Reynolds
stress associated with convection in the superadiabatic layers.
In stellar conditions the acoustic emission increases as $M_t^{7.5}$
where $M_t$ is the turbulent Mach number (Goldreich \& Kumar 1990).
Recent computations suggest that $M_t$ reaches a maximum around
$1.6 M_{\sun}$ on the zero age main sequence (ZAMS) 
and so does the oscillation amplitude (Houdek et al. 1999).
The most massive stars we address here are therefore expected to
have the largest fluctuation amplitudes.
In terms of surface velocity, the observations suggest :

\begin{equation}\label{eq1}
{\bf \upsilon_{osc} \propto (L_{\star}/M_{\star})^{0.7}}
\end{equation}

In this relation, as in the following ones,
$L_{\star}$ and $M_{\star}$ stand for stellar
bolometric luminosity and mass in solar units respectively.
Kjeldsen \& Bedding (1995) and Houdek et al. (1999)
have proposed similar laws with somewhat higher exponents.
However they predict large velocity fluctuations that 
exceed the actual observations in the case of most
massive stars such as Procyon ($1.46 M_{\sun}$).
Using theoretical models of stochastic excitation
(Samadi \& Goupil 2001), Samadi et al. (2004) confirm this
power-law dependence with an almost identical exponent (0.8 instead of 0.7).
Space-borne experiments will measure brightness
modulations into which $\upsilon_{osc}$ has to be converted.
Such an analysis is performed in Kjeldsen 
\& Bedding (1995) and we will not reproduce it here.
As the main result of the computations one obtains 
$\left(\frac{\delta L}{L}\right)_{Bol} \propto \frac{\upsilon_{osc}}{\sqrt{T_\mathrm{eff}}}$
which together with Eq.(\ref{eq1}) yields an estimate of the dependence 
of the amplitude of the luminosity
fluctuations on fundamental stellar parameters :

\begin{equation}\label{eq2}
\left(\frac{\delta L}{L}\right)_{Bol} \propto \frac{1}{\sqrt{T_\mathrm{eff}}} \left(\frac{L_{\star}}{M_{\star}}\right)^{0.7}
\end{equation}

If we now use the Sun as a calibrator, this simple law provides
an estimation of the luminosity variations induced by oscillations. Such an assumption 
should be adequate, for the stars we consider have solar-like
(convectively powered) oscillations and the external convection
properties should vary smoothly with stellar type. 
The oscillations associated with solar luminosity fluctuations are of the order
of 4 ppm (Kjeldsen \& Bedding 1995). Therefore if we express the stellar mass and luminosity
in solar units we finally deduce :

\begin{equation}\label{eq3}
\left(\frac{\delta L}{L}\right)_{Bol} = 4\times \sqrt{\frac{5780}{T_\mathrm{eff}}} \left(\frac{L_{\star}}{M_{\star}}\right)^{0.7}
\end{equation}

Using this last equation, Table \ref{tab1} gives
the amplitude of the luminosity fluctuations
as a function of mass for the Hyades models.
It also indicates the magnitudes of the corresponding
objects in the other two clusters.
Because of differences in composition and evolutionary stage,
the relative amplitudes of the fluctuations decrease slightly at a given
mass when going from Hyades to the younger clusters.
We do not compute any metallicity impact on the amplitudes.
The small variations are instead most probably due to an age effect
which is expected to be very small for $\sim 0.5$ Gyr difference
(Houdek et al. 1999) and we do not reproduce them here in detail.
The maximum variation concerns the most massive star where it
reaches 0.6 ppm i.e. $\approx$ 6 \%.

Also of interest is the range of frequencies for which
solar-like p-modes oscillations should be the best visible.
This frequency range roughly covers a region from the frequency $\nu_\mathrm{max}$ where
the largest amplitude is expected up to the cutoff
frequency $\nu_\mathrm{c}$ and a similar frequency extent below $\nu_\mathrm{max}$. 
We provide $\nu_\mathrm{max}$ in Table \ref{tab1} together
with the atmospheric cut-off frequency $\nu_\mathrm{c}$ ;
$\nu_\mathrm{c}$ is computed from its usual expression
$2\pi \nu_\mathrm{c}=\frac{c_s}{2H_\mathrm{p}}$ with $c_s$ and $H_\mathrm{p}$ respectively 
the sound speed and pressure scale height in the atmosphere ;
$\nu_\mathrm{max}$ is estimated following 
$\nu_\mathrm{max} = \frac{\upsilon_\mathrm{max}}{H_\mathrm{p}}$
where $\upsilon_\mathrm{max}$ stands for the maximum convective velocity
in the superadiabatic layers and $H_\mathrm{p}$ is the corresponding
pressure scale height. Because of the sharp drop in density and
temperature on the one hand and the requirement for nearly
constant convective flux on the other hand, convective motions
are substantially accelerated in the upper superadiabatic layers.
The driving of solar-like oscillations is highly 
sensitive to the kinetic energies of convective eddies. Thus
its efficacy reaches a maximum in the upper superadiabatic
layers. Modes of frequencies $\nu_\mathrm{max}$ similar to the local turn-over
frequency $\nu_\mathrm{to}$ where the convective velocity 
is maximal exhibit the largest oscillation amplitudes
and therefore $\nu_\mathrm{max} \approx \nu_\mathrm{to}$.
For modes having $\nu_\mathrm{osc} > \nu_\mathrm{to}$ the excitation vanishes because
of a lack of coherence between eddies at higher frequencies; 
while in contrast for $\nu_\mathrm{osc} < \nu_\mathrm{to}$ there is a rough
equipartition between convective kinetic and mode energies
(Goldreich \& Keeley 1977). However because of the increasing of the 
inertia of the modes the amplitudes diminish with the decrease of $\nu_\mathrm{osc}$.
We made a rough estimate
for the convective velocity $\upsilon$ using the mixing length theory
formulae (precise demonstrations are to be found in text books):

%\begin{equation}\label{eq4}
%{v = \frac{\alpha_\mathrm{MLT}}{2}\sqrt{\frac{1}{2}g\delta H_\mathrm{p}}\frac{\Gamma}{B^{1/2}}
%\vspace{1cm}
%B=\frac{1}{162}\frac{(\alpha H_\mathrm{p})^4 (\rho C_\mathrm{p})^2 \delta g}{K^2 H_\mathrm{p}}
%}
%\end{equation}

%\vspace{0.5cm}
\begin{eqnarray}\label{eq4}
\upsilon&=&\frac{\alpha_\mathrm{MLT}}{2}\sqrt{\frac{1}{2}g\delta H_\mathrm{p}}\frac{\Gamma}{B^{1/2}} \\
B&=&\frac{1}{162}\frac{(\alpha_\mathrm{MLT} H_\mathrm{p})^4 (\rho C_\mathrm{p})^2 \delta g}{K^2 H_\mathrm{p}} \nonumber 
\end{eqnarray}

\noindent
where the various symbols have their usual meanings :
$\alpha_\mathrm{MLT}$ is the mixing length parameter; g is the local
gravity; $\delta=-\left(\frac{d \ln \rho}{d \ln T}\right)_{P}$; $\Gamma$ is
the convective efficiency; $C_p$ is the gas specific heat at constant
pressure ; $\rho$ and $K$ are respectively
the gas density and thermal conductivity. 
These various quantities are derived from our models.
We use the Eq. \ref{eq4} as a scaling law considering
that in the case of the Sun the maximum oscillation amplitude is
at $3.3$ mHz which corresponds to the
well known 5-minute oscillations. Our solar reference
model for convection velocities is the calibrated 
'Btz' model from Brun, Turck-Chi\`eze \& Zahn (1999)
for which the equation \ref{eq4} provides a maximal
velocity $\upsilon_{max}=2.6\,\mathrm{km~s^{-1}}$.

The magnitudes and oscillation amplitudes computed
here are indicative of the observational accuracy 
needed to address solar objects in nearby open clusters using seismology.
Even though the COROT space mission does not primarily
aim at clusters we will consider its performance
as indicative of what is now within technical reach.
For a $m_V$=9 magnitude the threshold of oscillation
detection expected for this mission is
$\left(\frac{\delta L}{L}\right)_{Bol}=2.4$ ppm (Baglin et al. 2001).
Table \ref{tab1} therefore suggests that Hyades stars around and above 
solar mass
should become accessible to seismic
measurements in the near future. A Hyades star of $0.95 M_{\sun}$ is
for instance interesting in several respects.

It should exhibit $m_V$ in the 8.7 to
9.1 range depending on the actual age and composition of the Hyades
(see Table \ref{tab3} of Section \ref{sec3}). This somewhat
high magnitude is however compensated by a $\left(\frac{\delta L}{L}\right)_{Bol}$ we expect to be above 2.8 ppm.
In the next section we develop the informations on mass one could
extract from such a seismic target.
For the Pleiades and $\alpha$ Persei the situation is far 
less favourable because of their distances. With $1.3 M_{\sun}$ and $1.6 M_{\sun}$
stars having a visual magnitude of $\approx 9$
in these respective clusters, solar mass
stars will be difficult targets there.
The oscillation amplitudes expected above $1.3 M_{\sun}$ are sensibly
higher than solar ones and render corresponding stars better targets.
One nevertheless needs to be cautious concerning the granulation noise
which may limit the observations of luminosity fluctuations.
We end this section with two remarks:
1) Because of its youth a Hyades solar analog has a slightly
larger magnetic and chromospheric activity than our Sun (see 
Paulson et al. 2002 for recent evaluations). 
% typical Chromospheric log Rhk flux = -4.5 hyades of B-V=0.65
%  solar log Rhk flux = -4.9.
This could be problematic for the clear detection and 
and interpretation of the oscillatory modes.
2) Stars having masses between $1.5$ and $1.6 M_{\sun}$ are
$\delta$ Scuti variables when they are near to the ZAMS.
We are presently not interested
in the difficult question of low frequency modes excited by radiative
processes. In order to avoid such modes we shall limit ourselves
in subsequent computations to modes of orders above $n$=10 for which the $\kappa$
mechanism has been shown to be inefficient (Michel et al. 1999).
We note that both radiatively and convectively driven oscillations
are expected in $\delta$ Scuti variables. Furthermore because the
regions of $\delta$ Scuti involved in the driving
of the modes are not at dynamical equilibrium the expected luminosity fluctuations
$\frac{\delta L}{L}$ are much higher
than predicted when using equation (\ref{eq4})
(see Samadi et al. 2002).

\section{Global parameters and seismology in the Hyades}\label{sec3}

\subsection{Composition determination}\label{sec31}

Asteroseismology may complement spectroscopic chemical
determinations. It will provide some indications
on composition as, because of opacity effects,
metals and helium have a direct impact on stellar
global parameters (luminosity, effective temperature) and structure.
This information is of course dependent on the various
assumptions made in the modelling (convection parameter, diffusion processes, 
age...)\footnote{The past year's developments concerning solar surface oxygen fraction
however reminds us that spectroscopic measurements
also are strongly dependent on models (Asplund et al. 2004).}.
   \begin{table}
      \caption[]{The five composition considered for the Hyades.}
         \label{tab2}
     $$ 
         \begin{array}{p{0.4\linewidth}lllll}
       & C1 & C2 & C3 & C4 &C5 \\
      \hline
      \hline
	Helium mass fraction & 0.26 & 0.25 & 0.24 & 0.28 & 0.26 \\
      \hline
	$\mathrm\,[Fe/H]^{\mathrm{a}}$ (dex)   &  0.127 & 0.127 & 0.127 & 0.127 & 0.127 \\
%	\& iron group metals  &  &  &  &  & \\
      \hline
	$\mathrm\,[O/H]^{\mathrm{a}}$ (dex)  &  -0.07 & -0.07 & -0.07 & -0.07 & 0.127 \\
         \end{array}
     $$ 
\begin{list}{}{}
\item[$^{\mathrm{a}}$] $\mathrm [Fe/H]$ represents iron and 
iron-group metals while $\mathrm [O/H]$ represents oxygen and non-iron-group metals. 
\end{list}
   \end{table}

\begin{figure}[Ht]
\centering
\includegraphics[angle=90,width=9cm]{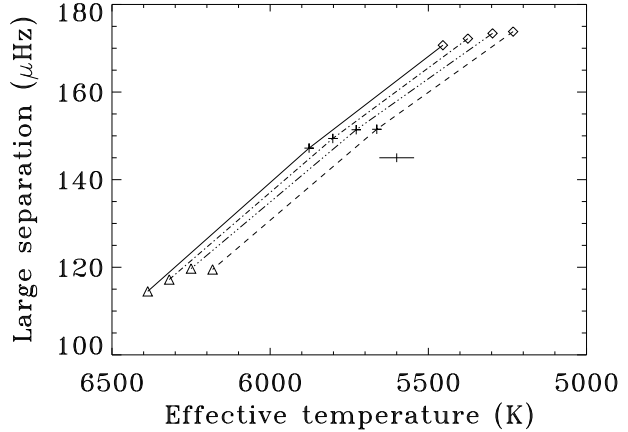}
\caption{Large separation as function of effective temperature
between $\sim$ 5000 and 6500 K for varying possible composition model of the 
Hyades. The solid line and dashed line correspond to
C1 ($[Fe/H] \neq [O/H], Y=0.26$) and C5 ($[Fe/H] = [O/H], Y=0.26$) 
respectively. For the same value of the abscissa they are
separated by $\sim 15\, \mu\mbox{Hz}$. The dot-dashed line and the triple dot-dashed
line represent C2 ($[Fe/H] \neq [O/H], Y=0.25$) and C3 
($[Fe/H] \neq [O/H], Y=0.24$) respectively. 
These models show that similarly to the usual HR diagram a decrease of the
helium fraction has the same effect as the increase of the metal fraction. 
The diamonds, crosses and triangles are the positions of the
0.95, 1.07 and 1.25 $M_{\sun}$ models for each composition.
The horizontal error bar below the curves at $\approx$ 5600 K
shows the typical uncertainty of the effective temperature. We
also indicated a $1\,\mu\mbox{Hz}$ wide vertical bar. The expected precision
for mode frequency determination from EDDINGTON is $0.3\,\mu\mbox{Hz}$.}
\label{fig1comp}
\end{figure}

As a first step, we will exploit
the collective seismic behaviours of stars 
of similar age and composition. In open clusters
it is indeed possible to draw conclusions from seismic HR diagrams
in the same way as from the classical HR diagram.
Figure \ref{fig1comp} is a transposition of the usual 
theoretical HR diagram where bolometric luminosity has been replaced
by the large separation. This figure therefore does not
show the traditional seismic HR diagram in which the 
small separation (defined as $\delta \nu_{\ell,n}=\nu_{\ell,n}-\nu_{\ell+2,n-1}$) 
is plotted as a function of the large 
separation (defined as $\Delta\nu_{\ell,n}= \nu_{\ell,n}-\nu_{\ell,n-1}$)
(Christensen-Dalsgaard 1988). 
In this study we have considered five plausible compositions for the Hyades,
summarized in Table \ref{tab2}.
The first composition (hereafter C1) assumes 
that $[Fe/H]=0.127$ dex (Boesgaard \& Friel 1990) applies to all 
iron-group metals we consider \footnote{Ni, Fe, Mn, Cr, Ti}
but uses $[O/H]=-0.07$ (Garcia Lopez et al. 1993) for oxygen and other metals.
It therefore presents a non-solar-fraction repartition among metals.
For this C1 composition we choose a helium fraction $Y=0.26$ as
supported by fitting of the lower main sequence (Perryman et al. 1998) and
by mass-luminosity relation of binaries (Lebreton et al. 2001).
The second (C2), third (C3), and fourth composition (C4) only differ 
from C1 through Y=0.25 Y=0.24 and Y=0.28 respectively. The fifth composition (C5) 
assumes $[Fe/H]=0.127$ dex, Y=0.26 and (the usual) solar repartition among metals.
Provided that the effective temperature is
reliably determined, Fig. \ref{fig1comp} illustrates that it should be possible 
to make a clear distinction between C1 and C5 :
in the 5000 to 6000 K effective temperature range the mean
large separation $\Delta \nu$ (see 
Sect. \ref{sec32} for the precise definition)
varies by $\sim 10 \mu\mbox{Hz}$ between 
a metal-rich Hyades star model (considering that
the metal fractions follow the [Fe/H]=0.13 dex value) and
its metal-poor counterpart (considering [O/H]=-0.07
dex as Garcia Lopez et al. (1993) suggests) that exhibits
the same effective temperature.
Such a difference is therefore quite significant when translated
to expected accurracies of seismic experiments.
The typical uncertainty in effective temperature 
found in the Hyades low-mass stars is 55 K (Thorburn et al. 1993)
whereas the large separation determination should be affected by 
an error around $0.1\,\mu\mbox{Hz}$.
The temperature uncertainties therefore have a much larger impact
on the interpretation of Fig. \ref{fig1comp} in terms of composition
of the Hyades.

\begin{figure}
 \begin{center}
 \rotatebox{90}{\includegraphics[width=4.8cm]{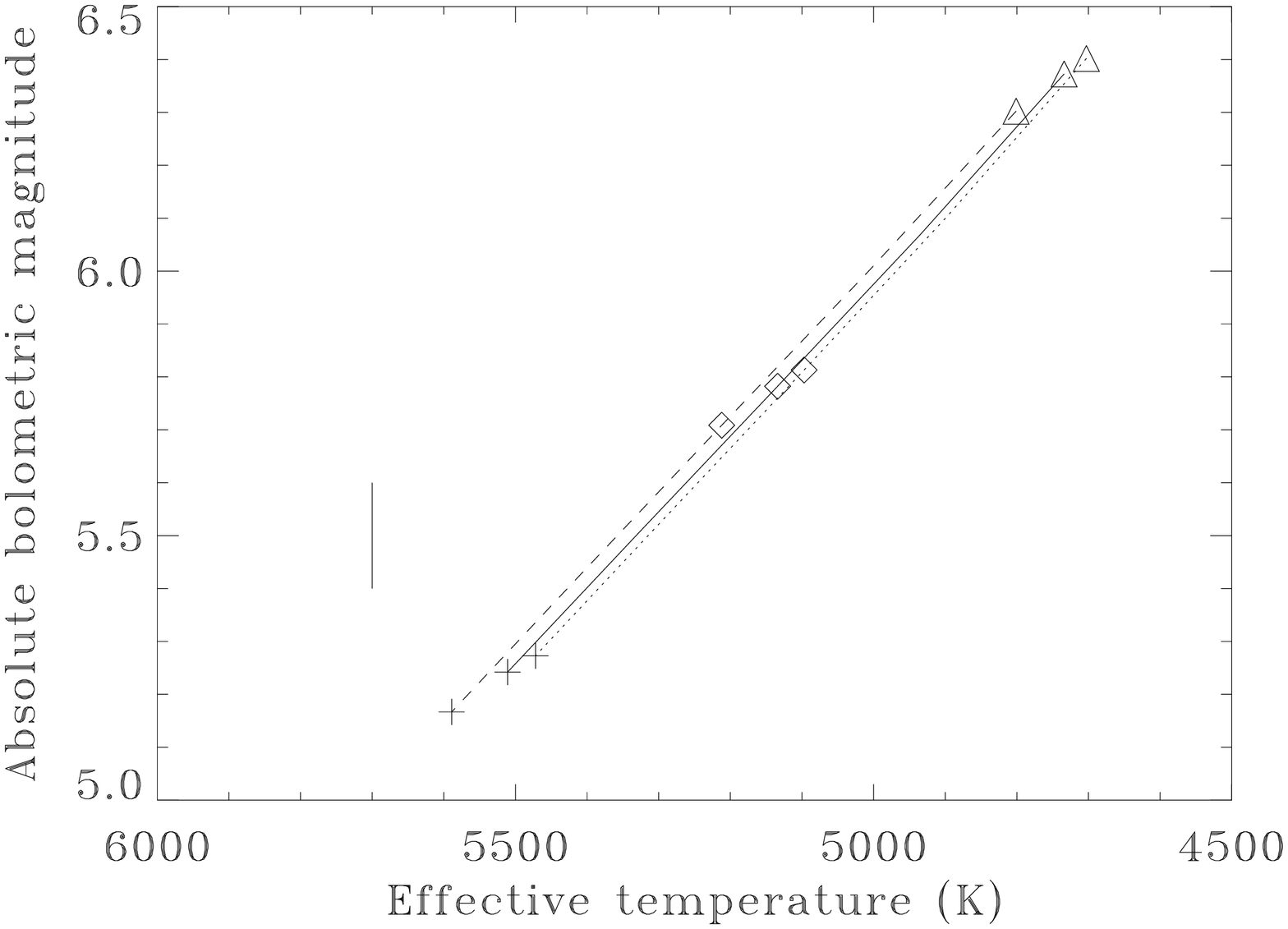}}
 \rotatebox{90}{\includegraphics[width=4.8cm]{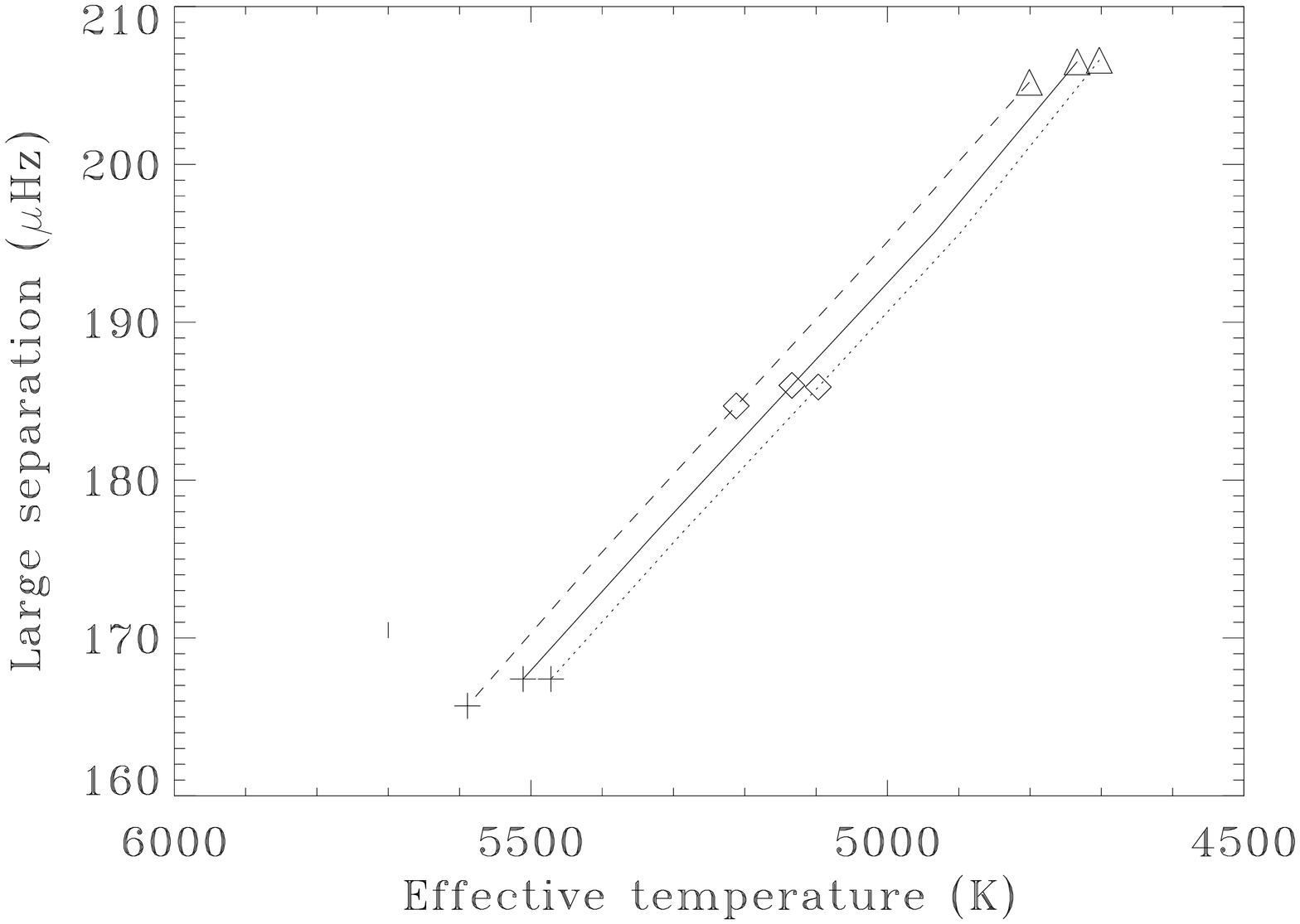}}
 \caption{Left panel: usual theoretical HR diagram for Pleiades.
Continuous line: solar composition
stars (X=0.7354, Y=0.2465); Dashed line: helium
fraction is increased by 0.1 but [Z/H] is kept unchanged so that X=0.7256,
Y=0.2565. Dotted line: [Z/H]=0.05 dex with respect to solar composition 
so that X=0.7275, Y=0.2524. The vertical error bar (0.2 mag) is representative of the
absolute magnitude uncertainty for the Pleiades cluster.
Triangles, diamonds, and crosses
show 0.8, 0.9 and 1 $M_{\sun}$ stars respectively.
Right panel: seismic HR diagram for the same cluster: Large separation vs effective
temperature. The plot conventions are similar. The minute vertical error bar of
$1 \mu\mbox{Hz}$ is representative of the maximal uncertainty in the large separation.
We note that the separation between main sequences is much larger
in the seismic diagram.}
 \label{fig2comp}
 \end{center}
\end{figure}

It is worth estimating the limits of this method.
Although the main sequences in Fig. \ref{fig1comp} strongly depend on composition
caution is needed when interpreting their positions in these terms.
The large separation is directly fixed by the radius and consequently
by the assumptions about convection modeling. For instance for a Hyades 
C1 composition and 0.95 $M_{\sun}$ model a change of $\alpha_{MLT}$ from our
standard value 1.76 to 1.56 induces decreases of 70 K and 4 $\mu\mbox{Hz}$
in $T_{eff}$ and mean $\Delta \nu$ respectively. These variations of convective
efficacy remain plausible (see Sect. \ref{sec4}). Let us now
compare the accuracy of the determination of the composition from
our seismic HR diagram with the 
precision of the classical HR diagram.
The smaller the uncertainties in the large separation (respectively the
luminosity) the better defined the main sequence and the more 
precise the composition determination through a fit of the lower
main sequence or the age determination from the turn-off position.
In the framework of usual HR diagram analysis, Stello \&
Nissen (2001) establish that the mean error in absolute visual magnitude can be
expressed as 
$\sigma(M_V) = \left[12 \sigma(b-y)^2 + \left(2.17 \frac{\sigma ( \pi )}{ \pi }\right)^2\right]^{1/2}$,
where b and y are the color bands from the Str\" omgren 
ubvy photometry while $\pi$ is the parallax.
The precision in $M_V$ for individual members of clusters
is mainly affected by uncertainties in these quantities.
In the case
of the Pleiades and the $\alpha$ Persei clusters the HIPPARCOS
value for $\frac{\sigma ( \pi )}{\pi}$
of the brightest stars exceeds 0.1 (Eggen 1998).
On the other hand Stello \& Nissen (2001) estimate
$\sigma (b-y) \approx 0.01$. The global error
in the absolute magnitude $M_V$ of individual stars  
in the Pleiades and $\alpha$ Persei
therefore is > $(0.12^2+0.21^2)^{1/2} \approx 0.25$.
In Fig. \ref{fig2comp} we illustrate and compare the impact of the composition
on positions of the main sequence in a classical and in the seismic HR diagram.
We consider three plausible compositions for the Pleiades
that present variations of $\Delta Y = 0.1$ and
$\Delta [Fe/H]=0.05$ dex among each other. We remark that with
respect to the errors the separation
along the ordinate axis is much larger in the
seismic than in the classical HR diagram.

%In the case of the Hyades the 1-1.5 mas accuracy measurement 
%in trigonometric parallaxes of individual stars
%achieved with HIPPARCOS translates into 0.1 mag precision on $M_V$
%(section 9 of Perryman et al. 1998). 
Recently de Bruijne et al. (2001)
derived extremely accurate individual
Hyades distances making use of HIPPARCOS secular 
parallaxes. This brings the accuracy in the absolute
magnitudes $M_V$ down to the $\sim 0.05$ level
and allowed the authors to produce an unprecedentedly narrow
main sequence. $M_V$ is however not a
quantity provided through stellar structure computation and 
that can be used in a theoretical HR diagram.
We note that the $B-V$ to $T_\mathrm{eff}$ calibration is still controversial
(Bessell et al. 1998, Lejeune et al. 1998) and is found to be inappropriate
below $\sim 5000$ K (de Bruijne et al. 2001).
%For such low temperatures the molecular
%opacity effects induced by the complex atmospheric
%chemistry ($TiO, H_2O, VO$) renders the $(B-V)-T_\mathrm{eff}$
%relation more and more hasardous.
Different transformations lead to zero age main
sequences varying by up to 0.2 mag in the $M_V, B-V$ plane and between
$B-V=0.7-0.8$ (ie $T_\mathrm{eff} \sim [5500,5200]$ K) whereas they are similar in the 
$M_{Bol},T_\mathrm{eff}$ plane (Robichon et al. 1999).
Moreover the $M_V$ transformation to bolometric luminosity is subject to
the difficult question of bolometric corrections ($BC_V$).
$BC_V$ depends on the effective temperature
and also critically depends on metallicity (Alonso et al. 1996)
and surface gravity (Arribas \& Martinez-Roger 1988).
We consequently stress here that as long as detailed metal
abundances remain uncertain absolute luminosity
determination might be biased.
In view of this different sources of uncertainty we
consider that seismic diagrams could give
valuable insight into composition even for the nearest
open cluster.

%\begin{figure}[H]
%\psfig{file=SHR32.ps, width=6cm, angle= 90}
%\caption{Absolute magnitude as function of acoustic radius. The masses
%are specified for the C1 composition with the same
%conventions as for figure \ref{fig1comp} : the cross,
%diamond and triangles are the positions of the
%0.8, 0.9 and 1 $M_{\sun}$ models. We added here a square and a cross for
%the 1.15 and 1.3 $M_{\sun}$ models. In the case of the solar
%mass stars (triangles) all composition models are indicated.
%Magnitude and acoustic radius vary so that the different models
%approximately remain on the same main sequence.}
%\label{fig2comp}
%\end{figure}

In Figs. \ref{fig1comp} and \ref{fig2comp} the main sequences 
are not surprisingly dependent on the helium fraction. Slight opposite 
variations in the helium or metal fractions have similar
impact on stellar luminosity and effective temperature so that a variation 
of the helium
fraction or metal fraction do not seem to be a priori
distinguishable. Hence main sequence fitting can only lead to metal 
abundances once the helium fraction is known (and conversely).
We nevertheless point out that if the Hyades metal fraction 
is around -0.07 dex instead of 0.1 dex
then the main sequence clearly lies above this last metal-rich 
main sequence for any possible helium fraction : the C3
composition main sequence is in a sense a limit that 
can be assigned to metal-poor main sequences for it
has Y=0.24 which is critically close to the actual primordial
helium evaluations : Izotov \& Thuan (1998) propose $Y_\mathrm{p}=0.245\pm0.004$
and Peimbert \& Peimbert (2000) propose $Y_\mathrm{p}=0.2345\pm0.003$.
%on the basis of HII regions observations in the Small Magellanic Cloud.

\begin{figure}
\centering
\includegraphics[angle=90,width=9cm]{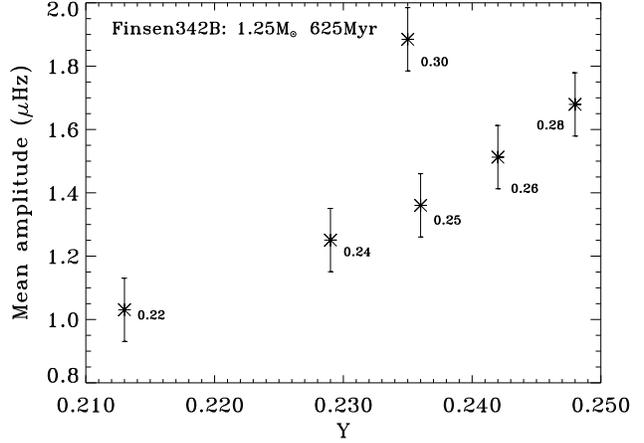}
\caption{The mean amplitudes of the second difference modulation
induced through the second ionization of helium 
scaled by $\Delta \nu / {\Delta\nu}_{\sun}$.
The means are given as function of helium fraction
in the envelope at the age of the Hyades : Y=0.213, 0.229, 0.236, 0.242, 
0.248, 0.235 and for composition C1 (crosses). Because of the diffusion process these
values 
do not correspond to the initial helium fractions that are indicated next
to the crosses.
Error bars of $0.1 \mathrm{\mu Hz}$ are only indicative.
%The means is also
%given for a C5 composition Y=0.26 (diamond).
}
\label{fig3comp}%seissHR5.pro
\end{figure}

In order to specifically check for the helium effects we have used 
the second difference : $\delta_2 \nu_{\ell,n}= \nu_{\ell,n-1} -2\nu_{\ell,n} + \nu_{\ell,n+1}$.
As is well known, the fluctuations of $\delta_2 \nu_{\ell,n}$ with frequency
outline the presence of two rapid variations in the stellar 
structure : firstly the region of the second ionization of helium, secondly
the  discontinuity at the base of the outer convection zone.
In a recent work Basu et al. (2004) suggest that it
is possible to determine the helium fraction Y using 
the mean amplitude of the second difference modulation
provided that both mass and radius are accurately known.
In that work the signature appears more clearly around 1.2 $M_{\sun}$
(see also Lopes et al. 1997).
Masses are generally not known in the Hyades and there are
no measurements of radii. However the age is known and some binary systems such
as Finsen 342 have their masses precisely determined. Because 
of its mass ($1.25 M_{\sun}$) and luminosity ($m_V \approx 7.4$) 
Finsen 342 B should be an interesting
target for helium determination.
We have built models of this star \footnote{Interestingly the $\theta^1$ Tau B star has
a very similar mass ($1.28 \pm 0.13 M_{\sun}$) to that of Finsen 342 B so that 
our analysis also applies to this object.}and computed the second
differences for different initial helium Y 
values. Assuming thar the second ionization of helium acts as a discontinuity,
the second difference was fitted to the form \footnote{The region of the 
second ionization of helium has a non-negligible extent and one should
use other fits than Eq. \ref{eq5}.
However these fits provide similar helium fraction estimations as the
one based on Eq. \ref{eq5}, see Basu et al. (2004)}:
%\begin{equation}\label{eq5}
\begin{eqnarray}\label{eq5} %JB
\delta_2\nu&=&\left(a_1+a_2\nu+\frac{a_3}{\nu^{2}}\right) \nonumber \\
&+&\left(b_1+\frac{b_2}{\nu^{2}}\right)\sin(4\pi \nu \tau_{He} +\phi_{He}) \nonumber \\
&+&\left(c_1+\frac{c_2}{\nu^{2}}\right)\sin(4\pi \nu \tau_{BCZ} +\phi_{BCZ})
\end{eqnarray}
%\end{equation}
Finally the amplitude $\bar A_{He}$ of the helium-induced modulation was
taken as the mean value of $b_1+\frac{b_2}{\nu^{2}}$ over
the the range $\mathrm{[1.5,3] mHz}$, both quantities being
scaled by $\Delta \nu / {\Delta\nu}_{\sun}$ (we take ${\Delta\nu}_{\sun}=135 \mu\mbox{Hz}$).
This last factor is introduced because the amplitude of the signal 
scales with the mean density, as does the large separation 
$\Delta \nu$ (see Basu et al. 2004).
Figure \ref{fig3comp} illustrates the results:
a variation
in helium fraction from 0.24 to 0.26 induces a change of $0.23 \mu\mathrm{Hz}$
in the fitted mean amplitude $\bar{A}_{He}$. Let us make here an important remark. 
In Hyades stars of $\approx 1.2 M_{\sun}$,
diffusion is definitely not a negligible process
because of the shallower outer convection zone.
In the models at 625 Myr, the initial outer 
convective region helium fractions of 0.24 and 0.26 have
decreased to 0.229 and 0.242 respectively. The
$\Delta \bar{A}_{He}$ variation is therefore larger than $0.1 \mu\mathrm{Hz}$ 
for $\Delta Y=0.01$. The diffusion effect becomes even more
dramatic above $Y=0.28$ as the extension of the convection zone
continuously diminishes with
increasing helium fraction. The model that has an initial helium fraction
of 0.3 indeed exhibits a lower Hyades age content ($Y=0.235$) in its convection
region as the model having initial helium fraction
of 0.28. In this instance we recall that the evolution of 
the $\bar{A}_{He}$ amplitude
is obviously not only related to the helium fraction but also depends
on physical conditions in the region of helium second ionization.

% Ancien emplacement des figures 1 et 2

\begin{figure}
\centering
\includegraphics[angle=90,width=9cm]{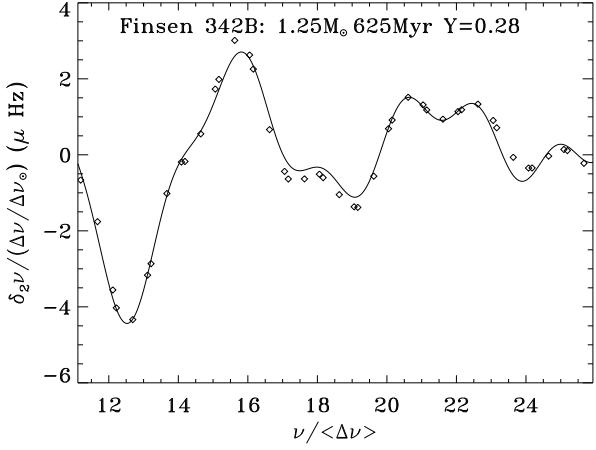}
\caption{Second difference and fit for Finsen342 B if a 
helium fraction of Y=0.28 is assumed.}
\label{fig4comp}% sndiffhelaurent.pro
\end{figure}

\subsection{Mass determination}\label{sec32}

Binary systems have been extensively used to compute stellar masses
on the basis of astrometric and spectroscopic measurements. 
A recent work of S\"{o}derhjehm (1999) based on Hipparcos astrometric data evaluates 
individual masses within 27 systems. For the subsample of
14 systems within
the approximate mass range from 0.5 to 1.5 $M_{\sun}$
the mean precision in mass determination is 0.11  $M_{\sun}$
and exceeds 0.06 for all of the 
stars but two. In the Hyades cluster no system has its
mass determined to better than $0.16\,M_{\sun}$ and individual masses are
available for only one system with a precision better than $0.2\,M_{\sun}$. 
Making simultaneous use of spectroscopy and speckle interferometry
can significantly improve the accuracy of such mass determinations : 
Pourbaix (2000) determines the masses to better
than 5 percent for 25 stars from 40 binary systems. In very favourable cases 
the mass uncertainty has been decreased to slightly below
$0.01 M_{\sun}$ (Latham et al. 1996).
Combining observations of the Hyades binary system
Finsen 342 (70 Tauri) over a period of 15 yr, Torres,
Stefanik \& Latham (1997B) found $1.363 \pm 0.073 \, M_{\sun}$ 
and $1.253 \pm 0.075 \, M_{\sun}$ for the primary and the secondary components.
However, because of the observational difficulties and the long
amount of time required for careful interferometric and spectroscopic measurements
only very few other stars in the Hyades (vB22, 51 Tau, $\theta^{2}$ Tau) 
have had their masses
determined with similar precision (Torres Stefanik \& Latham (1997A);
Peterson \& Solensky 1988).

Asteroseismology opens up new possibilities for accurate 
absolute or relative mass determinations. Its methods 
are not restricted to close binary systems but extend
the determination of masses to all stars. We note by the way that the non-occurrence 
of oscillations could be used as an indication of possibly 
unresolved multiple systems that are sometimes
suspected (S\"{o}derhjelm 1999) in classical observations.
In the following we focus on Hyades stars in the 0.77 to
1 $M_{\sun}$ mass range. 

\begin{table}[ht]
  \begin{center}
    \caption[]{Stellar parameters for 0.769(vB22B),
0.8, 0.95 and 1 $M_{\sun}$ Hyades stars of composition C1. For $0.95 M_{\sun}$ results are given for various ages and compositions. Large 
separation computations are based on the mean of $\Delta \nu_{\ell,n}$ in the frequency range going 
$3500$ to $4500$ $\mu\mbox{Hz}$.
The scatter in this mean always lies between 0.5 and 0.7 $\mu\mbox{Hz}$. 
Only low degree modes l=0,1 and 2 have been taken into account
in the computations.}\vspace{1em}
    \renewcommand{\arraystretch}{1.2}
    \begin{tabular}[h]{lccccc}
      	Composition & Mass         &  Age  & $T_\mathrm{eff}$ & $m_v$ & $\Delta\nu\ $   \\
      	            & ($M_{\sun}$) & (Myr) & (K)              &       & $(\mu\mbox{Hz})$ \\
      \hline
	C1 & 0.769 & 625 & 4707 & 10.19 & 206.2 \\
      \hline
	C1 &  0.8  & 625 & 4835 & 9.92  & 199.1  \\
      \hline
	C1 &  0.95 & 625 & 5454 & 8.84  & 170.2  \\
      \hline	      	     
	C2 &  0.95 & 625 & 5375 & 8.93  & 171.5  \\
      \hline	      	     
	C3 &  0.95 & 625 & 5297 & 9.02  & 172.7  \\
      \hline	      	     
	C4 &  0.95 & 625 & 5612 & 8.70  & 167.0  \\
      \hline	      	     
	C5 &  0.95 & 625 & 5232 & 9.10  & 173.1  \\
      \hline
	C1 &  0.95 & 500 & 5451 & 8.84  & 170.8  \\
      \hline	      	     
	C1 &  0.95 & 700 & 5456 & 8.83  & 169.8  \\
      \hline
	C1 &  1.00 & 625 & 5637 & 8.54  & 160.8  \\
      \hline
      \end{tabular}
   \label{tab3}
  \end{center}
\end{table}

The asymptotic theory of p-mode frequencies (Tassoul 1980) 
makes it possible to relate
the large separation $\Delta\nu_{\ell,n}= \nu_{\ell,n}-\nu_{\ell,n-1}$ to the 
stellar acoustic radius $\tau_R$, namely :

\begin{equation}\label{eq6}
\frac{1}{2 \Delta\nu_{\ell,n}} \sim \int_{0}^{R} \frac{dr}{c} = \tau_R
\end{equation}

\begin{figure}
\centering
\includegraphics[angle=90,width=9cm]{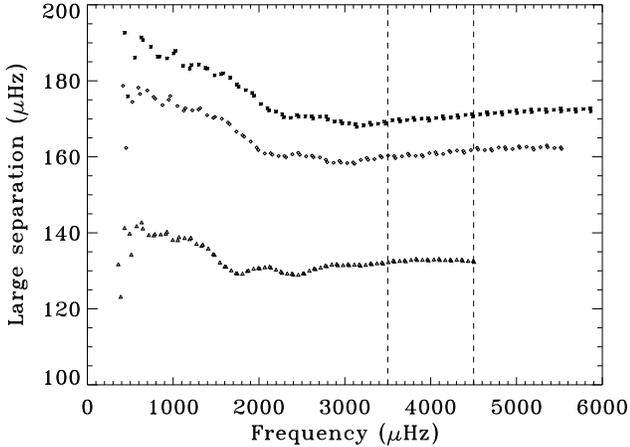}
\caption{Large separation as function of frequency and for different masses;
crosses : 0.95 $M_{\sun}$, diamonds : 1. $M_{\sun}$ and triangles : 1.15$M_{\sun}$
The vertical dashed lines shows the 3500-4500 $\mu\mbox{Hz}$ range over which
we consider the mean large separation.}
\label{fig1mass}% seissdiffrnces1.pro
\end{figure}

Table \ref{tab3} shows the mean $\Delta\nu$ for masses
varying from $\sim 0.77$ to $1 M_{\sun}$ 
and for possible ages and compositions of the Hyades.
We compute $\Delta\nu$ as the average 
of $\Delta\nu_{\ell,n}$ over $\ell=0,1$ and $2$ and for frequencies between 3.5 and 4.5 mHz.
The lower limit is approximately the frequency of the mode of 
radial order n=15 and degree $\ell=0$ in a $0.77 M_{\sun}$ model.
For higher mass stars this frequency corresponds to 
higher orders because of the decrease of the acoustic radius with mass. It
therefore always lies above n=10 which keeps the explored domain
far from possible radiatively induced oscillations
(see Sect. \ref{sec2}). The other reason why we utilize
frequencies in this range is that, as can be seen in Fig. \ref{fig1mass},
the large separation reaches the asymptotic regime
above $\approx$ 3 mHz; thus its direct interpretation in terms of acoustic radius
becomes all the more reliable. Finally we set an upper limit of
4.5 mHz on the computation of the mean large 
separation in order to reduce the scatter in this mean.
We preferentially select the models at 625 Myr 
which has recently be estimated to be the Hyades age
(Perryman et al. 1998). However we also extend the analysis to 500 and
700 Myr to check for the impact of evolution slightly outside the claimed 
age uncertainties for the cluster $\approx 50$ Myr. 
Calculations of the large separation of lower mass stars ($0.95 M_{\sun}$) show its 
value to be extremely accurate. In the frequency range considered
the large separations are extremely flat; the scatter is
$\sim 0.5\mu\mbox{Hz}$ (Fig. \ref{fig1mass}). Moreover we find that the large separation
experiences very small variations with age. 
The very slow evolution of stars having masses below a solar mass explains 
this stability. The $0.9M_{\sun}$ star main sequence lifetime
exceeds 14 Gyr (Schaller et al. 1992). We note that the
large separation is also very stable 
with composition (for any conceivable variation of the metal or helium fraction).
We report a $\Delta\nu$ variation of $\sim 5.7 \mu\mbox{Hz}$ ie $3.4\%$
of its total value for $\Delta Y $ as high as 0.04. 
The different values for the helium fraction determined by
Pinsonneault et al. (1998)
(Y=0.283) and Perryman et al. (1998) (Y=0.26)
lead to a $\Delta \nu$ variation of $3.2 \mu\mbox{Hz}$. 
The variation with metallicity is similarly small.
Going from composition C1 to C5 there 
is roughly 0.2 dex difference in metal content 
(iron group metals represent less than $8\%$ 
of the global metal mass fraction). Then we note however that 
$\Delta \nu$ only changes by $2.9 \mu\mbox{Hz}$.
It is well known that metals mostly affect the stellar
plasma opacity. The opacity in turn changes the stellar
stratification in radiative zones but has a much weaker effect in outer convection
zones where stratification is mostly adiabatic. Yet this is precisely the part
of the star that mainly contributes to a large separation value.
As illustrated by the aforementioned asymptotic expression
the large separation is determined where the sound speed 
is at its lowest, i.e. in the coolest regions.
In the $0.95 M_{\sun}$ star of C1 composition the convection
zone represents $0.29 \, R_{\star}$ but conversely its acoustic
radius is $0.59$ \% of the total acoustic radius. 
For the other compositions we check that these values are 
different by less than 3 \%.

The age- and composition-related stability of $\Delta \nu$ make it
easier for the effect of mass to show up.
A change of $0.05\,M_{\sun}$ gives rise to a $\approx 10\mu\mbox{Hz}$ 
discrepancy. This variation clearly is 
much larger than the variations associated with mass or composition, 
and also larger than the resolution
level of future space-borne experiments
such as EDDINGTON that should determine
frequencies to the level of $0.3 \mu\mbox{Hz}$ (Favata et al. 2000).
Hence simple measurements of the large separation variation
might be used for the mass determination of lower main sequence single 
Hyades stars. The accuracy in mass determination will reach a few 
percent of solar mass. As oscillations are expected for any 
solar-like star the sample of these objects 
having accurately determined masses will be greatly increased. 

The predicted mass accuracy is comparable to or above th usual precision
achieved when one relies on binary systems. It improves
towards lower mass objects where the age determination uncertainties are negligible.
However, as the mass decreases some other difficulties arise
because of higher complexity
in the input equation of state or 
low-temperature opacities (Chabrier \& Baraffe 1997).
Besides this many aspects of the physics of convection in the outer
regions of cool stars ($T_\mathrm{eff}=4000 - 5000$ K)
remain to be investigated (Ludwig et al. 1999; 2002).
A straightforward interpretation of 
the large separation in terms of mass might seem optimistic and it 
is perharps safer to use the large separation as an indicator of the 
mass hierarchy. By chance the vB22 system provides the best known
mass in the Hyades : its secondary component ($0.769 \pm 0.005 M_{\sun}$; 
Peterson \& Solensky 1988) lies in the region we have studied (Table \ref{tab3}).

\section{The extension of outer convection in the Hyades}\label{sec4}

The precise extension of convective mixing in stars is crucial in several respects.
First it has a direct impact on stellar evolution.
Namely the occurrence of lower overshooting would significantly affect the
evolution of low mass objects after they leave the main sequence and when they
ascend the red giant branch (Alongi et al. 1991).
The history of the surface abundances of the light elements
closely depends on the extent of the convection zone.
This in turn strongly constrains any inference on the dynamical 
phenomena in envelopes (Piau \& Turck-Chi\`eze 2002; Piau et al. 2003). 
Finally convection and rotation 
remain the key ingredients for improving our
knowledge of the angular momentum evolution or of the 
surface magnetic activity.
In this instance observational studies of young stars are now capable
of providing new clues on the precise nature of the dynamo process
responsible for surface magnetic manifestations (Feigelson et al. 2003;
Donati et al. 2003).

Thanks to the increase of computational power, direct 
three-dimensional modelling increasingly approches
realistic physical conditions of stellar convection.
Many of these studies have focused
on solar-type outer convection zones (Brummell et al. 2002 and references therein)
and now also begin to address convective cores (Browning et al. 2004).
However, the corresponding convective flows are extremely swift and turbulent.
Consequently it is not possible to model convection in its full
complexity
and a fortiori to model stellar evolution in the framework of such direct
approaches. For almost half a century we have relied on
parametrized theories for convection such as the
mixing length theory (B\"{o}hm-Vitense 1958).
As is well known the MLT possesses one main free parameter $\alpha_\mathrm{MLT}$ that
is empirically adjusted by means of measurements of the actual solar 
radius. Some studies of close binary systems suggest
that no change of $\alpha_\mathrm{MLT}$ is required from the Sun to
other low mass stars (Fernandes et al. 1998) while others
suggest some slight changes $\approx 0.2$ with mass, e.g. in the $\alpha$ Cen system 
(Eggenberger et al. 2004). 
There are no basic physical reasons why
this parameter should be independent of age, mass or composition.

\begin{figure}
\centering
\includegraphics[angle=90,width=9cm]{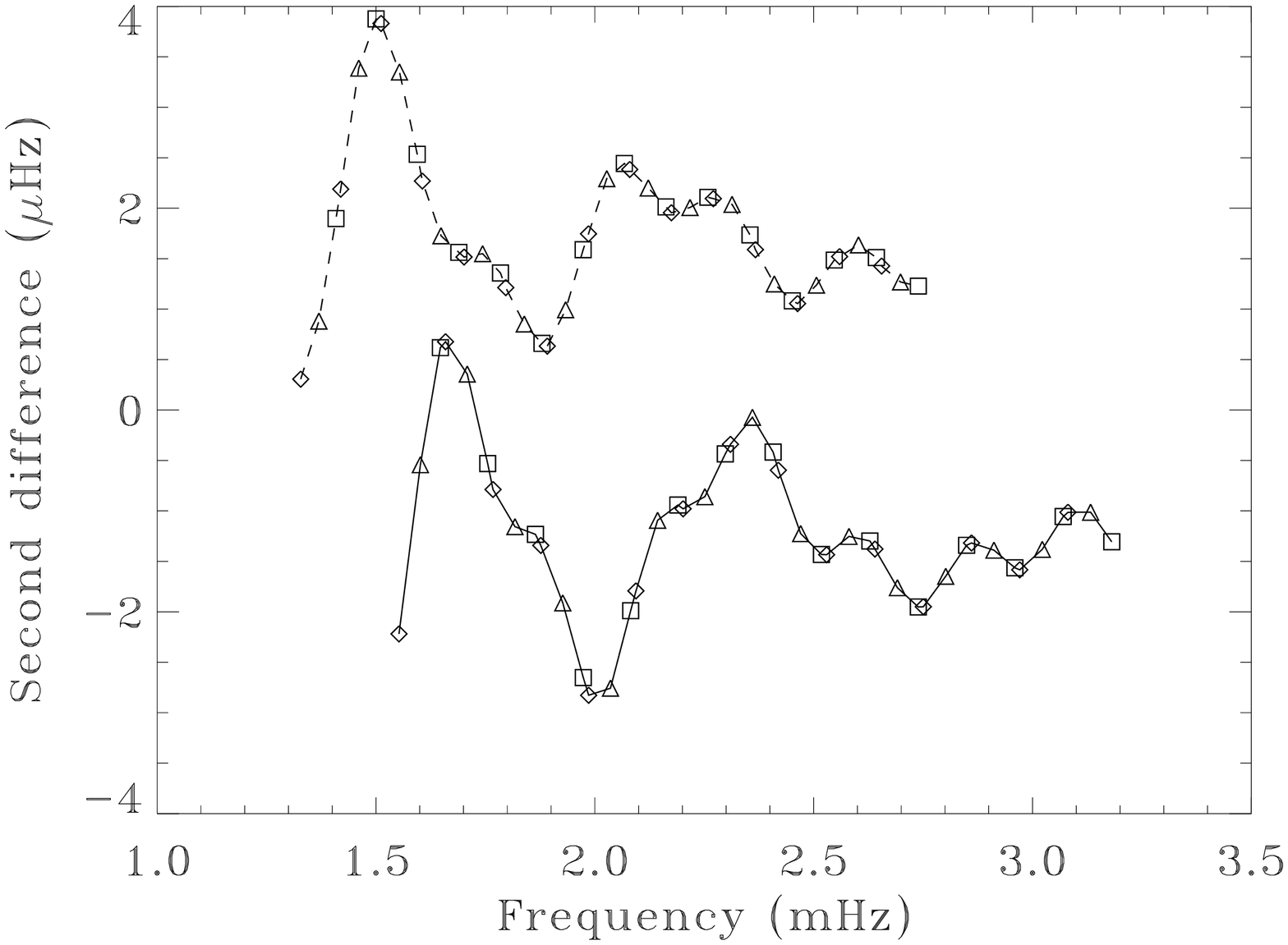}
\includegraphics[angle=90,width=9cm]{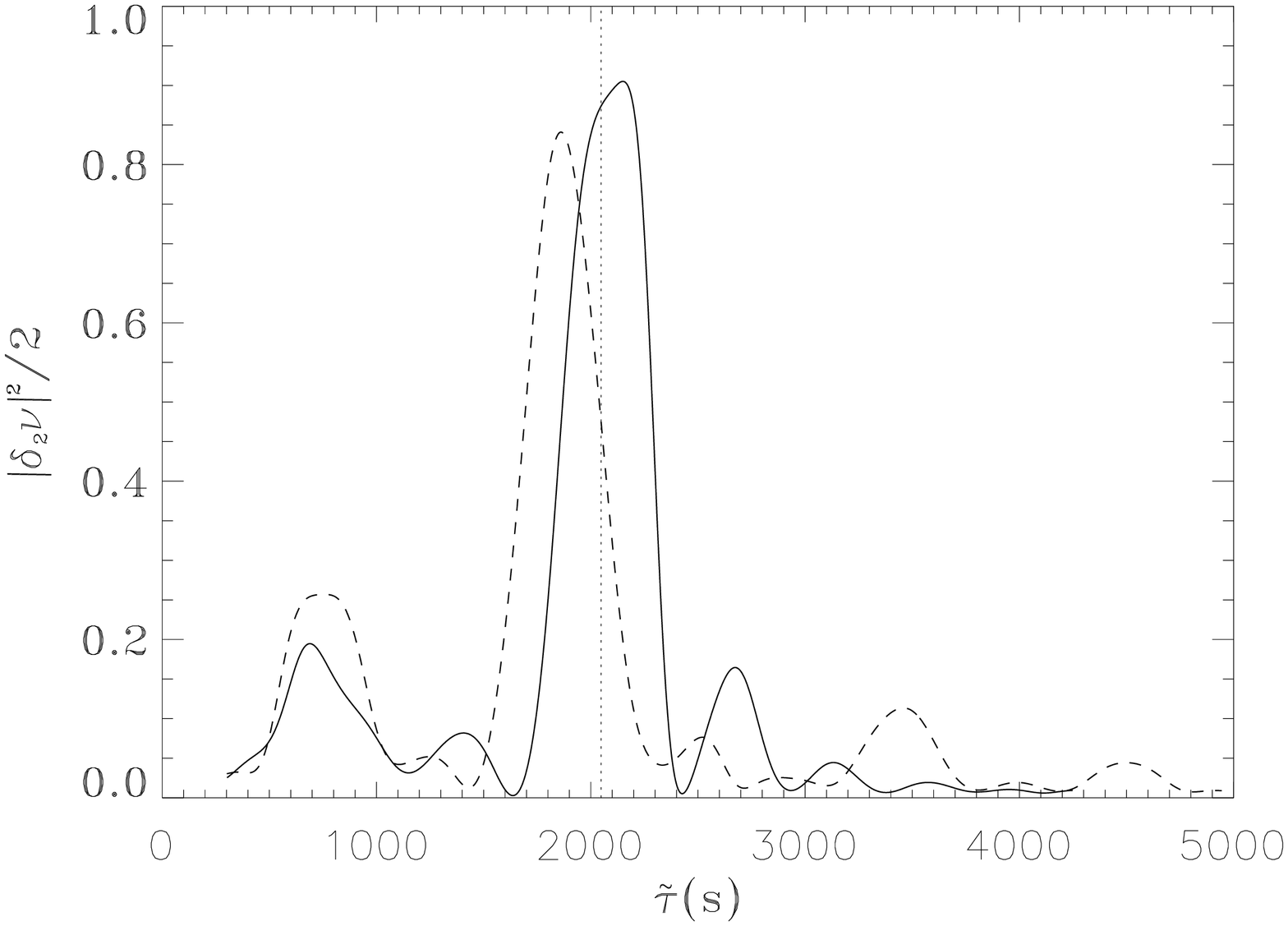}
\caption{Upper panel : the second difference as function of frequency for 
two Hyades models (625 Myr) of similar $T_\mathrm{eff} \approx$ 6550 K 
but with different $\alpha_\mathrm{MLT}$ and masses.
$\diamond : \ell=0$, $\triangle : \ell=1$, $\square : \ell=2$.
Dashed line : 1.35 $M_{\sun}$ and $\alpha_\mathrm{MLT}=1.566$.
Continuous line : 1.3 $M_{\sun}$ and $\alpha_\mathrm{MLT}=1.966$.
In order to make the figure clearer the upper and lower tracks
have been shifted by 1.5 and -1.5 $\mu\mbox{Hz}$ respectively along the ordinate
axis with respect to the original values.
Lower panel : amplitudes for the 
Fourier transforms of second differences presented
in the upper panel with similar conventions on linestyle.
A smooth interpolation of second 
differences has been performed and subtracted to raw data before 
the Fourier transform in order to attenuate the He II 
ionization effect (this effect is still visible through the 
peaks around 800 s).
The vertical dashed line shows the middle of the continuous
line peak at half maximum.}
%lit_freq_pbdouble.pro & plotspectres.pro
\label{fig1astruct}
\end{figure}

Our goal here is to check how seismic information
can be used to acquire better knowledge of convection.
On the one hand the convective efficiency
in the superadiabatic layers directly affects the extent of the convection zone.
On the other hand seismology has provided direct measurements
of the depth of the convection zone in the
solar case (Christensen-Dalsgaard et al. 1991).
It is therefore possible to relate seismic measurements to
convective efficacy.
Thanks to recently developed techniques
one can address the issue of the depth of the outer convection zone
in other stars than our Sun (e.g. Monteiro et al. 2000; 
Roxburgh \& Vorontsov 2001; Ballot et al. 2004, hereafter BTG04): the presence of a 
discontinuity in stellar structure is detectable through (low-degree) mode phase 
shifts where it induces oscillatory behaviour. For instance
the second difference ($\delta_2 \nu_{\ell,n}= \nu_{\ell,n-1} -2\nu_{\ell,n} + \nu_{\ell,n+1}$)
exhibits modulations whose periodicities
are directly related to the acoustic depth of the discontinuities
(see Fig. \ref{fig1astruct}).
We quantitatively evaluate the variations of this parameter
resulting from the convection / radiation discontinuity 
for various Hyades models in the 1 to 1.3 $M_{\sun}$
range. This range lies above the masses addressed
in binary systems by Fernandes et al. (1998) and is interesting
to address in the Hyades in several respects.
As open clusters comprise stars with
marginally different compositions or ages, they offer
the possibility to specifically check the impact of mass on the convective efficiency 
in the superadiabatic layers. 
Then the oscillations will be more easily detected above $\approx 1 M_{\sun}$ as the corresponding
stars exihibit visual magnitudes below $\approx 8 m_{V}$ and show larger fluctuations
than the Sun and their lower mass counterparts (cf Sect. \ref{sec2}). 
At the same time a Hyades member of $\approx 1 M_{\sun}$ should have lost
a significant fraction of its initial
angular momentum and its rotation rate should be small all the way 
down to its center (Piau et al. 2003).
This is an important point as the present analysis does not
take into account rotationally induced effects on models and
oscillation frequencies. In this instance
Hyades observations suggest a rapid decrease of the equatorial
velocity as a function of $T_\mathrm{eff}$ between 7000 and 6000~K.
As suggested by Gaig\'e (1993) whereas below $T_\mathrm{eff}=6400$~K there are no stars with
an equatorial velocity above $30 \,\mathrm{km~s^{-1}}$ the stars above $T_\mathrm{eff}=7000$ K 
exhibit a distribution with a large scatter 
from 0 up to $\approx 200 \,\mathrm{km~s^{-1}}$.
For masses of 1, 1.15 and 1.3 $M_{\sun}$ we compute
$T_\mathrm{eff}=5640, 6110, 6500$ K  respectively at the 
age of the Hyades and the mass interval should
therefore correspond to slow or moderate rotators.
%La question du lithium gap et de la diffusion
Finally the mass interval considered covers the red wing of the so called
lithium gap (Boesgaard \& Tripicco 1986): stars with the same age as the Hyades or older 
having effective temperatures between 6300 and 6850 K are lithium- and
also beryllium-deficient in comparison to hotter or
cooler stars. This is most probably related to a
rotationally induced (magneto)hydrodynamic mixing process 
occurring at the top of the radiation region.
For almost 20 years the lithium gap has been an active research field
both theoretically (Talon \& Charbonnel 1998)
and observationally (Boesgaard \& King 2002).

To determine the extent of the convection zone
we make use of the second difference $\delta_2 \nu_{\ell,n}$,
exclusively evaluated from modes of
degree $\ell=0, 1$ and $2$. For the order $n$ we have considered 
the range between 13 and 26 which corresponds to the most excited modes
in the solar case. 
%The second difference is exploited directly
%ie we make no evaluation of the noise that will affect real data.
Two periods are clearly seen in $\delta_2 \nu_{\ell,n}$ vs $\nu_{\ell,n}$
(see Fig. \ref{fig1astruct}): the smallest oscillation is 
associated to the base of the convection zone (hereafter BCZ).
The largest corresponds to the discontinuity
induced by the second ionization region of helium which
has been discussed in Sect. \ref{sec31}.
To reduce its impact we subtract 
a smoothed interpolation of the raw data from $\delta_2 \nu_{\ell,n}$. 
This procedure is more precisely described in BTG04.
We then perform a spectral analysis to evaluate the acoustic depth
of the convection zone $\tau_{BCZ}$ (Fig. \ref{fig1astruct}). 
The transform gives $\tilde{\tau}_{BCZ}$ which is an estimate
of $\tau_{BCZ}$ biased by surface effects (Christensen-Dalsgaard et
al. 1995). In the solar case there is for instance
a 200 s difference between $\tau_{BCZ}$ and $\tilde{\tau}_{BCZ}$.
Similarly half the inverse
of the large separation, $\frac{1}{2\Delta \nu}$, is 
shifted by this amount with 
respect to the acoustic radius $\tau_R$ (cf equation \ref{eq6}). The difference 
$\frac{1}{2\Delta \nu}-\tilde{\tau}_{BCZ}$ is therefore unbiased and 
corresponds to the acoustic radius of the convective base $\Delta \tau_{BCZ}$ (BTG04).

\begin{equation}\label{eq7}
\Delta \tau_{BCZ}=\tau_R-\tau_{BCZ} \approxeq \frac{1}{2\Delta \nu}-\tilde{\tau}_{BCZ}
\end{equation}

\begin{figure}
\centering
\includegraphics[angle=90,width=8cm]{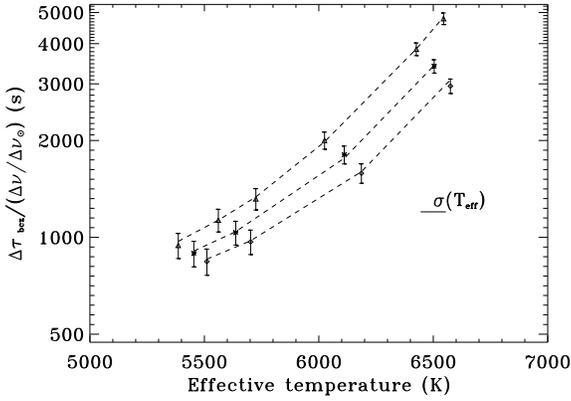}
\caption{Acoustic radii of the bases of the outer convection zones 
scaled by $\Delta \nu / {\Delta\nu}_{\sun}$
and as a function of the effective temperature at the age of the Hyades. 
Three $\alpha_\mathrm{MLT}$ have been considered.The symbols show results 
extracted from $\delta_2\nu$.
Triangles : $\alpha_\mathrm{MLT}$=1.566, masses 0.95, 1, 1.05, 1.15,
1.3 and 1.35 $M_{\sun}$; Stars : $\alpha_\mathrm{MLT}$=1.766, 
masses 0.95, 1, 1.15 and 1.3 $M_{\sun}$.
Diamonds : $\alpha_\mathrm{MLT}$=1.966, masses 0.95 , 1, 1.15 and 1.3 $M_{\sun}$.
The dashed lines indicate the true BCZ positions in the different models
The horizontal error bar $\sigma (T_\mathrm{eff})$ is $2 \times 55$ K.
The vertical error bars are discussed in the text.
}
%alphahyades.pro
\label{fig2struct}
\end{figure}

We use $\Delta \tau_{BCZ}$ as it is not (or less) dependent than $\tilde{\tau}_{BCZ}$ 
on surface effects
where the physical description is less under control
(turbulent pressure, superadiatic effects).
Figure \ref{fig2struct} shows the acoustic radii of convective bases $\Delta \tau_{BCZ}$ 
scaled by $\Delta \nu / \Delta \nu_{\sun}$
as a function of effective temperatures and for the three $\alpha_\mathrm{MLT}$ values
1.566, 1.766 (our solar calibrated value)
and 1.966. The acoustic radii scales with the mean
density and inversely with frequencies. As in Fig.
\ref{fig3comp}, the scaling factor $\Delta \nu / {\Delta\nu}_{\sun}$
therefore makes it possible to isolate the effect of the $\alpha_{MLT}$ parameter. 
It moreover reinforces our ability to determine changes in $\Delta \tau_{BCZ}$
with effective temperature because it extends the variation range of $\Delta \tau_{BCZ}$.
The higher $\alpha_\mathrm{MLT}$ higher the
convective efficiency and deeper the convection
zone. As observed on Fig. \ref{fig2struct} this in
turn translates into smaller acoustic
radii for the internal radiative regions.
On Fig. \ref{fig2struct} we
plot a typical $\sigma(T_\mathrm{eff})$ error bar of 55 K for
the effective temperature as estimated by Thorburn et al. (1993). 
%As $\Delta \tau_{BCZ}$ error bars we take the width at half maximum of the
%corresponding peaks in the spectral analysis
%(figure \ref{fig1astruct}).
The errors made in the determination of
$\Delta \tau_{BCZ}$ are extensively discussed in
BTG04. We summarize the results here.
Firstly there is a intrinsic error due to the method itself. This
includes the effects of the limited number of points and the unknown
residual bias of the estimator. For the modelled solar-like stars, this error
is smaller than 1.5\,$\%$, and generally smaller than 1\,$\%$ in units of
total stellar acoustic radius. This difference can be observed in 
Fig. \ref{fig2struct}
by comparing the symbol positions with respect to the dashed lines.
Secondly the effects of
the uncertainties on the frequencies can be estimated with Monte Carlo simulations.
For expected errors of 0.1--0.15~$\mu\mbox{Hz}$ on the frequencies,
the error induced in the determination of 
$\Delta \tau_{BCZ}$ is around 2--2.5\,$\%$.
Thus we have considered a pessimistic global error of 3\,$\%$ in units of
total stellar acoustic radius.

Using a 2D hydrodynamic code including radiative
transfer Ludwig et al. (1999) evaluate the stratification 
of the outer convection zone
for various surface conditions in $T_\mathrm{eff}$ and $\log\,g$.
They cover stellar types from K to F
($4300 \mathrm{K} \leq T_\mathrm{eff} \leq
7100$ K and $2.54 \leq \log\,g \leq 4.74$).
The $\alpha_\mathrm{MLT}$ should vary from $\approx 1.3$ to $\approx 1.75$
to match the entropy jumps they find. 

The 0.4 range we consider here for $\alpha_\mathrm{MLT}$ 
is therefore representative of the variations 
with surface conditions expected in convection efficiency.
Interestingly Lebreton et al. (2001) find that Hyades 
isochrones having $\alpha_\mathrm{MLT}=1.6-1.8$
(respectively $\alpha_\mathrm{MLT} \la 1.4$) better fit the 
$M_V-B-V$ observed magnitude
diagram in the $1.3-1.7 M_{\sun}$ range (respectively below $1 M_{\sun}$).
We note here that owing to the errors 
any variation of convective efficiency
with stellar type corresponding to an amplitude 
$\Delta \alpha_\mathrm{MLT} \approx 0.2$ should be clearly 
observable in a diagram similar to Fig \ref{fig2struct}.

To push the investigation further we have computed
stellar models with variable $\alpha_\mathrm{MLT}$ parameters.
Following the Ludwig et al. (1999) scaling-law, $\alpha_\mathrm{MLT}$ varies 
with the surface conditions ($T_\mathrm{eff}$ and
$\log\,g$). With regard to constant (solar-calibrated) $\alpha_\mathrm{MLT}=1.76$ 
models the difference in $\Delta \tau_\mathrm{BCZ}$ for
the variable $\alpha_\mathrm{MLT}$ models is quite small ($\la 20$~s) 
except for the $1.3 M_{\sun}$ star where it reaches 95 s. 
This stems from Ludwig's result that around solar conditions 
the $\alpha_\mathrm{MLT}$ values exhibit a plateau in the $T_\mathrm{eff}$ -- $\log\,g$ 
plane. The slightly larger  $\Delta \tau_\mathrm{BCZ}$
($\approx 100$~s) of the `variable` 
$\alpha_\mathrm{MLT}$ $1.3 M_{\sun}$ star reflects the decrease
in convective efficiency towards higher $T_\mathrm{eff}$ predicted by
Ludwig et al. (1999). Let us stress in this example the ability of seismology to
give constraints for the hydrodynamic convection
simulations (Robinson et al. 2004).

Table \ref{tab4} summarizes some quantities characterising the convection
together with the corresponding stellar surface conditions ($T_\mathrm{eff}$ and $\log\,g$). 
We recall here that although
$\alpha_\mathrm{MLT}$ surely is the most important parameter within the
MLT framework it is by no means the only one. There are indeed 
several formulations of the MLT that give different
stratification for identical $\alpha_\mathrm{MLT}$ values as 
illustrated by Ludwig et al. (2002). To specify the MLT formulation
we use, we provide four other parameters (such as the surface to volume
ratio for convective elements) in a short appendix.
Like any other local approach of convection (eg Canuto \& Mazzitelli 1991),
the MLT is simplistic and can be criticized on many points. Indeed
various $\alpha_\mathrm{MLT}$ parameters are required to match hydrodynamic computations 
for velocity or thermal stratification in the upper convection region (Ludwig et al. 2002).
For these reasons, we provide
a more meaningful physical quantity than $\alpha_\mathrm{MLT}$ :
the specific entropy of the deep convection zone $s_{cz}$. 
By entropy of the deep convection zone we mean the entropy
level reached where convective motions become extremely 
close to adiabatic. $s_{cz}$ is directly
related to the extent of the convection zone and to the convective
efficiency in superadiabatic layers. It may therefore
be assigned an intermediate role between hydrodynamic computations and seismology.
\begin{table*}[ht]
  \begin{center}
    \caption{Column 1 : mass. Columns 2 \& 3 : surface 
effective temperature $T_\mathrm{eff}$ and gravity $\log\,g$. Column 4 : $\alpha_\mathrm{MLT}$, 
column 5 : acoustic radius of the base of the outer
convection zone $\Delta \tau_{BCZ}$ as deduced from the Fourier transform of 
the second difference and typical error bars;
column 6 : specific entropy in the deep convective region $s_{cz}$.
The models constructed with a variable $\alpha_\mathrm{MLT}$
are to be found in the last section of the table. The reported $\alpha_\mathrm{MLT}$ 
values are those corresponding to the surface conditions at the age of the Hyades.}\vspace{1em}
    \renewcommand{\arraystretch}{1.2}
    \begin{tabular}[h]{lccccc}
      \hline
      	Mass & $T_\mathrm{eff}$ & $\log\,g$ & $\alpha_\mathrm{MLT}$ & $\Delta \tau_{BCZ}$ & 
      $s_{cz}$ \\
      	($M_{\sun}$) & (K) &  &  & (s) & ($\mathrm{erg~K^{-1}~g^{-1} / 10^9}$) \\
      \hline
	0.95 & 5386 &  4.54 &  1.566 & 1155$\pm$122   & 1.67 \\

	1.   & 5561 &  4.51 &  1.566 & 1304$\pm$120   & 1.72 \\

	1.05 & 5725 &  4.48 &  1.566 & 1426$\pm$152   & 1.77 \\

	1.15 & 6025 &  4.41 &  1.566 & 1892$\pm$157   & 1.88 \\

	1.3  & 6426 &  4.31 &  1.566 & 2934$\pm$198   & 2.19 \\

	1.35 & 6545 &  4.28 &  1.566 & 3394$\pm$194   & 2.38 \\
      \hline
	0.95 & 5454 &  4.56 &  1.766 & 1123$\pm$121   & 1.66 \\

	1.   & 5637 &  4.53 &  1.766 & 1232$\pm$118   & 1.70 \\

	1.15 & 6111 &  4.44 &  1.766 & 1777$\pm$165   & 1.85 \\

	1.3  & 6504 &  4.34 &  1.766 & 2697$\pm$206   & 2.12 \\
      \hline
	0.95 & 5511 &  4.57 &  1.966 & 1085$\pm$120   & 1.65 \\

	1.   & 5702 &  4.55 &  1.966 & 1185$\pm$119   & 1.69 \\

	1.15 & 6186 &  4.46 &  1.966 & 1609$\pm$130   & 1.83 \\

	1.3  & 6575 &  4.35 &  1.966 & 2419$\pm$216   & 2.39 \\
      \hline
	0.95 & 5458 &  4.56 &  1.781 & 1120$\pm$121   & 1.66 \\

	1.   & 5640 &  4.53 &  1.774 & 1209$\pm$130   & 1.70 \\

	1.05 & 5807 &  4.50 &  1.774 & 1387$\pm$142   & 1.74 \\

	1.15 & 6101 &  4.43 &  1.741 & 1769$\pm$164   & 1.86 \\

	1.3  & 6465 &  4.32 &  1.662 & 2792$\pm$193   & 2.15 \\
      \hline
      \end{tabular}
   \label{tab4}
  \end{center}
\end{table*}

We shall end this section with two caveats. Firstly and as 
noticed in the introduction no overshoot was considered here. 
Monteiro et al. (2000) show that
such a phenomenon would increase the 
amplitude and the acoustic depth of the signal 
resulting from the BCZ discontinuity.
Thus any interpretation of the extent of the outer convection zone
in terms of convective efficiency or $\alpha_\mathrm{MLT}$ parameter
depends on the overshoot. We nevertheless note that 1) the actual
helioseismic inferences on overshooting suggest that this phenomenon
is moderate, Monteiro et al. (1994) find $l_{ov}=0.07 H_\mathrm{p}$; 2) hydrodynamic
simulations of the solar convection zone similarly suggest rather
moderate overshooting : below $0.11 H_\mathrm{p}$ (Brummell et al. 2002).
The predicted movements moreover do not mix
the stable region with enough efficiency to change its stratification significantly
\footnote{they represent therefore strictly speaking overshoot and 
not penetrative convection.}
i.e. leave no seismic signature.
Secondly the diffusion process has been taken into account
following Michaud \& Proffitt (1993). Although we do not report 
it here in details we mention that the amplitude of the BCZ-induced oscillatory
behaviour in $\delta_2 \nu_{\ell,n}$ increases 
both with age and mass of our Hyades models. This is opposite to the comparable 
effects on frequencies reported in Monteiro et al. (2000).
Microscopic diffusion is responsible for this difference.
Monteiro et al. mostly addressed ZAMS stars where this process
has no impact and was therefore not included in their computations.
The description of diffusion we use 
is valid for G dwarfs and later types
but becomes unrealistic for earlier types.
Above $T_\mathrm{eff} \approx$ 6500 K the radiative diffusivity which we do not consider 
plays a non-negligible role (Morel \& Thevenin 2002). Besides this the
stellar rotation increases and the microscopic diffusion interacts more strongly 
with one or several
macroscopic mixing processes as it is suggested by the occurrence
of the lithium dip.

\section{Summary and conclusion}\label{sec5}

We computed classical and seismic
stellar parameters using the stellar evolution code CESAM and the
adiabatic pulsation package of Christensen-Dalsgaard.
Classical and seismic variables
were employed to constraint the stellar models.
Only $\ell=0,1,2$ low-degree modes were considered.
As a preliminary analysis we estimated apparent visual magnitudes,
luminosity fluctuations and a few other interesting seismic variables
in three nearby open clusters. We subsequently focused on stars within
the closest and best-known of these : the Hyades.
We predicted (Sect. \ref{sec2}) that the Hyades stars with masses
around or above solar should exhibit oscillations
detectable by the near-future seismic space-borne 
experiments. Future ground-based
observations with HARPS or ESPaDOnS, in probing surface velocities
could similarly begin to answer some of the questions mentioned here.
Exploiting the predicted oscillation frequencies we had three goals:
to set constraints on mass, composition and structure.
Let us briefly sum up our results on these points and conclude.

The mean value of the large separation $\Delta \nu$ in the
3.5 to 4.5 mHz frequency range is a strong function of the mass: we
calculated (Sect. \ref{sec32}) that a $\approx 0.05 M_{\sun}$ mass
change induces a $\approx 10 \mu\mbox{Hz}$ variation in $\Delta \nu$.
This is much higher than the $\Delta \nu$ variation expected from 
typical uncertainties in age or
composition (Table \ref{tab3}). Asteroseismology
will therefore allow the determination of absolute
and relative masses
in nearby open cluster and with competitive accuracy when
compared to the best astrometric and spectrometric mass evaluations to date.
However, at variance with these measurements seismology requires 
only short observation runs and is not limited to a very
small sample of binaries.

Asteroseismology can shed new light on the
composition of open clusters(Sect. \ref{sec31}). As main opacity contributors,
metals strongly influence the effective temperature $T_{eff}$
and have a direct impact on the internal structure which
affects the large separation $\Delta \nu$. Using Hyades
models of different masses we have shown that the location
of the $T_{eff}-\Delta \nu$ relation in the corresponding
diagram provides a measurement of the global metal content.
In particular, we are able to distinguish  models buildt 
with solar-intermetallic ratios
from models buildt with non-solar-intermetallic ratios. This is 
an important
issue as we recall that 1) iron only represents $\approx 2$ \%
of the total metal content in number 2) there is a clear scatter
in [O/Fe] vs [Fe/H] in the Galactic disk (Edvardsson et al 1993).
As is well known, the amplitude $\bar{A}_{He}$ of the
second difference $\delta_2 \nu$ main modulation stems
from the second ionization region of helium. Its use to infer
helium fraction in the envelope however requires other stellar
parameters such as mass and radius to be known (Basu et al. 2004).
The FINSEN 342 star belongs to a Hyades binary system. Its
mass and age are accurately known. Given these constraints, we 
showed that for this star a variation of the initial helium
mass fraction from 0.26 to 0.28 would increase $\bar{A}_{He}$
by $0.2 \mu\mbox{Hz}$.

The main feature of structure of solar-like stars is the 
presence of a radiative core surrounded by a convective region.
Using $\delta_2 \nu$ modulations with frequency we have
evaluated the acoustic radii of the bases of the convection zones (BCZ)
for 1 to 1.3 $M_{\sun}$ Hyades models and to an accuracy
better than 3\,$\%$ of the total acoustic radius (Sect. \ref{sec4}).
As, for a given effective temperature, the acoustic radius of the BCZ
depends on convective efficiency, asteroseismology
provides a possible tool for constraining
$\alpha_\mathrm{MLT}$ (Fig. \ref{fig2struct}).
As is well known, the mixing-length theory does not
give a physically consistent picture of convection. Nevertheless,
it is interesting to constrain $\alpha_\mathrm{MLT}$ in stars as the
ongoing hydrodynamic computations on convection make predictions
for this parameter. Seismology therefore offers a means
to check our calculations (and understanding) of convection
in the superadiabatic layers. The considered mass range
is interesting in several respects : it starts where the
knowledge of our Sun can presumably be transposed and then extends in a 
domain where variations in convection efficiency are
suspected through HR diagram analysis (Lebreton et al. 2001)
or binary systems calibrations (Eggenberger et al. 2004).
Moreover the interval covers the red wing of the lithium dip
where additional mixing is at work in the radiation zone
(Talon \& Charbonnel 1998). Setting constraints on convection
depth in this range will not only shed new light on the
convection efficacy in related stars but will also help constraining
any deeper dynamical mixing phenomenon.

Because they comprise stars of similar age and composition,
open cluster are unique laboratories for stellar physics.
In this respect the Hyades have proved to be of primary importance.
This importance is transposed to the asteroseismology field.
In the Hyades it is possible to take advantage of collective
effects and also to select some especially interesting
binaries. In this work we presented various 
seismic tests that could be performed assuming today's
accuracy on the measurement of oscillation frequencies. These tests clearly
open up new opportunities in the field of stellar physics.
Applied to the Hyades, asteroseismology will allow precise
mass and composition determinations which are crucial
for valuable astrophysical tests. It will furthermore
enable measurements of the depths of outer convection zones which in turn will 
provide indications of convective efficiency in stars.

\acknowledgements{
Laurent Piau is grateful to Reza Samadi, Marie-Jo Goupil, Ilidio Lopes and Dimitri
Pourbaix for their help and for useful discussions.
Laurent Piau is presently scientific collaborator to the Belgian F.N.R.S.
}

\appendix
\section*{Appendix}

The MLT version we use relies on the following equations (Cox \& Giuli 1968) 
where $f_1$, $f_2$, $f_3$ and $f_4$ are dimensionless free parameters 
each having a specific physical meaning :

$\upsilon^2 = f_1 \frac{\Lambda^2 g \delta ( \bigtriangledown - \bigtriangledown_e )}{H_\mathrm{p}}$

$\upsilon$ is the convection velocity.
$f_1$ describes the dissipation of the work of buoyancy forces due to viscosity and we
set it to $\frac{1}{8}$.

$F = f_2 \frac{ \rho c_\mathrm{p} \upsilon T \Lambda ( \bigtriangledown - \bigtriangledown_e )}{H_\mathrm{p}}$

$F$ is the convective energy flux.
$f_2$ is related to the proportion between upward and downward matter movements
and is set to $\frac{1}{2}$.

$\Gamma = \frac{\bigtriangledown - \bigtriangledown_e}{\bigtriangledown_e - \bigtriangledown_{ad}} = \frac{\rho c_\mathrm{p} \upsilon \tau_e}{f_3 \sigma T^3} (1+\frac{f_4}{\tau_e^2 })$

$\Gamma$ is the convection efficiency.
$f_3$ is related to the $surface / \Lambda \ volume$ ratio of convective elements
and is set to 24. Finally $f_4$ is related to radiative transfer
inside convective bubbles and is set to 3. 
Most of the symbols in the preceding equations keep their usual meanings 
(eg $\Lambda= \alpha_\mathrm{MLT} H_\mathrm{p}$ is the mixing length) and were
defined in Sect. \ref{sec2}. In addition : T is the temperature;
$\sigma$ is the Stefan-Boltzmann constant; $\bigtriangledown$, $\bigtriangledown_e$ and
$\bigtriangledown_{ad}$ are the mean thermal gradient, 
the convective element thermal gradient and the
adiabatic thermal gradient respectively; $\tau_e$ is the Rosseland optical thickness of a convective 
element. The four dimensionless parameters are similar to the B\"{o}hm-Vitense (1958)
formalism with the exception of $f_\mathrm{4}$ which she assessed to be zero.
As remarked by Henyey et al. (1965), however, the value of this parameter
only has a marginal impact on the general description of convection through
MLT. We have moreover directly checked that our approach and results indeed are similar
to other computations done within the B\"{o}hm-Vitense formalism.

\end{document}